%% file: ArXiv.tex
\documentclass[jidm,a4paper]{jidm} 
\usepackage{graphicx,url}  
\usepackage[T1]{fontenc}   
\usepackage[utf8]{inputenc}

\newdef{definition}[theorem]{Definition}
\newdef{remark}[theorem]{Remark}

\jidmVolume{}
\jidmNumber{}
\jidmYear{}
\jidmMonth{}
\include{DEFINE}

\title{Overcoming Bias in Community Detection Evaluation}

\author{Jeancarlo C. Leão, Alberto H. F. Laender and Pedro O. S. {Vaz de Melo}}

\institute{Instituto Federal do Norte de Minas Gerais (IFNMG), Brazil \\ \email{jeancarlo.leao@ifnmg.edu.br}
\and Universidade Federal de Minas Gerais, Brazil \\ \email{\{laender,olmo\}@dcc.ufmg.br}
}

\begin{abstract}
\input{abstract}
\end{abstract}

\category{H.2}{Data Mining}{Miscellaneous} 
\category{G.2.2}{Graph Theory}{Miscellaneous}
\category{J.4}{Social and Behavioral Sciences}{Miscellaneous}

\keywords{Community Structure, Quality Evaluation, Bias, Ensemble Approach, Triangulation Method}

\begin{document}

\begin{bottomstuff}
\end{bottomstuff}

\maketitle

\input{JIDMFinal}
	
\bibliographystyle{jidm}
\bibliography{bibliographydoi}

\begin{received}
\end{received}

\end{document}

%% file: DEFINE.tex
\usepackage[utf8]{inputenc}
\usepackage[nomessages]{fp}
\usepackage{graphicx,url}
\usepackage{booktabs}

\usepackage{sidecap}

\usepackage{makecell}

\usepackage{amssymb}
\usepackage{cancel}

\usepackage{url}
\usepackage{siunitx}
\usepackage[flushleft]{threeparttable}
\usepackage{multirow}
\usepackage{makecell}
\usepackage{array,xstring}
\usepackage{numprint}

\usepackage{verbatim}

\usepackage{array, tabularx, boldline}
\usepackage[inline]{enumitem}

\usepackage[table,xcdraw]{xcolor}

\newcommand{\prg}[1]{\vspace{6pt}\noindent{\textit{#1}}}

\newcommand{\citep}[1]{\cite{{#1}}}

\renewcommand{\theenumii}{\theenumi.\arabic{enumii}.}
\newcommand{\REVIEWED}[1]{{\textcolor[rgb]{0.2,0,0}{{#1}}}}

\newcommand{\iAPS}{{APS}}

\newcommand{\iPUBMED}{{PubMed}}

\newcommand{\nocitedb}[1]{}

\definecolor{darkgreen}{rgb}{0.0, 0.2, 0.13}
\definecolor{red}{rgb}{1, 0, 0}
\definecolor{max}{rgb}{0, 0, 0}
\definecolor{min}{rgb}{0, 0, 0}

%% file: abstract.tex
Community detection is a key task to further understand the function and the structure of complex networks. 
Therefore, a strategy used to assess this task must be able to avoid biased and incorrect results that might invalidate further analyses or applications that rely on such communities. 
Two widely used strategies to assess this task are generally known as \textit{structural} and \textit{functional}. 
The structural strategy basically consists in detecting and assessing such communities by using multiple methods and structural metrics. 
On the other hand, the functional strategy might be used when ground truth data are available to assess the detected communities. However, the evaluation of communities based on such strategies is usually done in experimental configurations that are largely susceptible to biases, a situation that is inherent to algorithms, metrics and network data used in this task. 
Furthermore, such strategies are not systematically combined in a way that allows for the identification and mitigation of bias in the algorithms, metrics or network data to converge into more consistent results. In this context, the main contribution of this article is an approach that supports a robust quality evaluation when detecting communities in real-world networks. In our approach, we measure the quality of a community by applying the structural and functional strategies, and the combination of both, to obtain different pieces of evidence. Then, we consider the divergences and the consensus among the pieces of evidence to identify and overcome possible sources of bias in community detection algorithms, evaluation metrics, and network data.
Experiments conducted with several real and synthetic networks provided results that show the effectiveness of our approach to obtain more consistent conclusions about the quality of the detected communities.

%% file: JIDMFinal.tex
\section{Introduction}

{The community detection problem has been much studied in the context of social networks due to its wide application in many domains, giving rise to many methods to address it~\cite{almeida2012towards,Fortunato201075,10.1140/epjds/s13688-020-00223-0,YangLeskovec2015}.}
However, one of the major challenges related to this problem is the difficulty to evaluate the detected communities with respect to the various methods proposed in the literature. Part of this difficulty lies on the fact that there is still  no universally accepted definition for the concept of community~\cite{Fortunato201075}, as well as for what we understand as being the quality of a community~\cite{Hric2014}. Besides, the evaluation of such communities is usually carried out by using experimental configurations greatly susceptible to biases, which are inherent to the algorithms, metrics and network data used in this task~\cite{JEBABLI2018651,LLD2019SBBD,Liu20158693534}. Moreover, this kind of evaluation is, in general, carried~out without explicitly dealing with such biases, which may lead to inconsistent results.

\begin{figure*}[t]
     \begin{center}            
\includegraphics[width=0.85\textwidth]{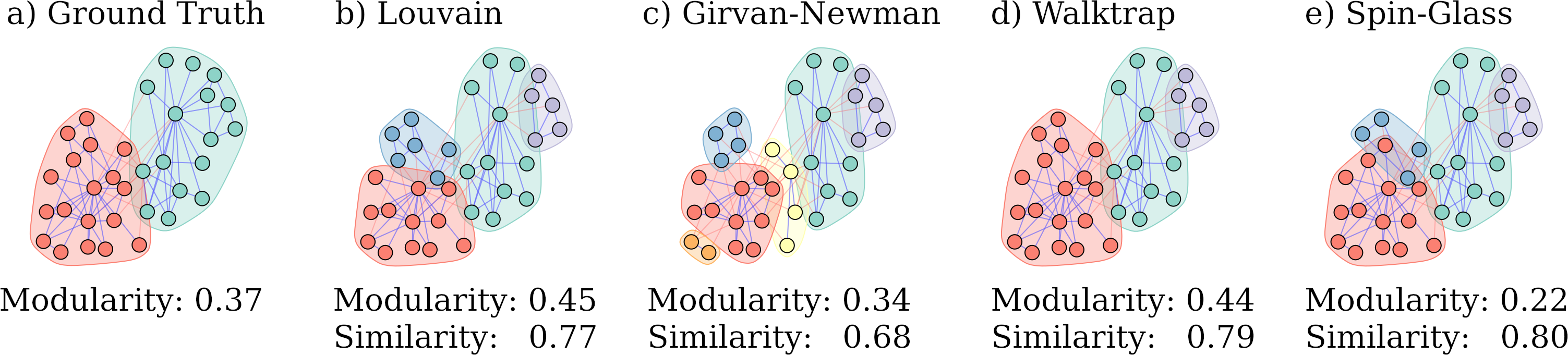}
    \end{center}
    \caption{Example of how bias can affect community  detection in social networks.}
    \label{graphic:ToyNetwork.pdf}
\end{figure*}

In order to illustrate this problem, let us consider the example shown in Figure~\ref{graphic:ToyNetwork.pdf}. Specifically, Figure~\ref{graphic:ToyNetwork.pdf}a shows a social network formed by 34 members (vertices) of a karate club interconnected by edges representing interactions between them outside the club. Originally, this network was divided into two non-overlapping communities labeled by Zachary~[1977]\nocite{zachary1977information} with 16 and 18 members, respectively, each one supervised by a specific instructor. Figure~\ref{graphic:ToyNetwork.pdf}b, on the other hand, shows the communities detected in this same network by the Louvain algorithm~\cite{blondel20081742-5468-2008-10-P10008}, a well known and very effective community detection algorithm. Note that the community structure revealed by the Louvain algorithm is different from those presented in Figures~\ref{graphic:ToyNetwork.pdf}c-e, 
which were respectively obtained by the Girvan-Newman~\cite{NewmanGirvanPhysRevE.69.026113}, Walktrap~\cite{Pons2005} and Spin-Glass~\cite{PhysRevE.74.016110} algorithms, all of them also considered very effective. Here, we raise the possibility that the bias of each heuristic algorithm interferes with its final results, making them different from each other and from the absolute optimal theoretical result, not necessarily present in Figure~\ref{graphic:ToyNetwork.pdf}.

Thus, let us check the pieces of evidence present in this example in order to reach a consensus on which algorithm produces the best quality communities. First, the modularity metric indicates that the communities shown in Figure~\ref{graphic:ToyNetwork.pdf}b present a better quality with respect to their modular structure (i.e., they present the highest modularity value). On the other hand, when comparing the detected communities with the ground truth (Figure \ref{graphic:ToyNetwork.pdf}a) using the Rand Index similarity metric~\cite{Rand-JASA1971}, it indicates that the network in Figure~\ref{graphic:ToyNetwork.pdf}d is the best one, since its communities are among the most modular ones, being also more similar to those shown in Figure~\ref{graphic:ToyNetwork.pdf}a, even though there is not a perfect match. 
However, due to its own bias this similarity metric scored better the communities in Figure~\ref{graphic:ToyNetwork.pdf}e, even though they show more visual differences with respect to the ground truth than the communities in Figure~\ref{graphic:ToyNetwork.pdf}d. Finally, due to some specific bias in the original network data or in the ground truth data, the Girvan-Newman algorithm has not been able to identify good communities (Figure~\ref{graphic:ToyNetwork.pdf}c), as shown by the values of the two metrics considered. 
Note that, although the modularity value of these communities is the one closest to the ground truth's, these two sets of communities are less structurally similar, since the number of subgraphs obtained by this algorithm is the largest among all networks. 

In face of these pieces of evidence pointing to opposite directions with respect to the quality of the communities in our example, a question arises on which one presents the best structure and what causes this divergence. We could try to obtain a consensual result, but this would be inadequate, because hypotheses about bias were not tested. In addition, evidence obtained from visual inspection is only viable on very small networks such as the ones considered in Figure~\ref{graphic:ToyNetwork.pdf}. 
Nevertheless, our assumption is that all pieces of evidence provide a significant amount of information about the network structure that, combined with other pieces of evidence on the main sources of bias, results in a consistent decision on the quality of the detected communities. 

To the best of our knowledge, there is no comprehensive evaluation approach that, considering multiple strategies, is able to identify which one provides the best interpretation. More importantly, as we detail later, due to their own biases it is not always possible to find a consensus among different metrics and community detection algorithms on which community structure has the highest quality. This requires a cross-checking approach involving at least three distinct evaluation strategies to indicate a consensus and estimate a possible bias with respect to the quality of the revealed communities. 

A strategy generally employed to improve methods, measures, and data reliability and validity is triangulation\footnote{According to O'Donoghue and Punch~[2003]\nocite{ODonoghue&Punch2003}, triangulation is a ``method of cross-checking data from multiple sources to search for regularities in the research data.''}, which consists in using different approaches for measuring the same characteristic~\cite{BRENDER2006253}. 
Note that, with only one measure of a specific characteristic, the error and biases inherent in that measure are confounded with the characteristic itself. Thus, when measuring different aspects using different metrics, it is possible to decide which one can bring into better focus the characteristic of interest. 
Based on this idea, the main contribution of this article is a robust approach for community quality evaluation that allows one to obtain results less prone to bias when detecting communities in synthetic and real networks. 

Thus, given a network, its set of ground truth communities and a set of its communities to be evaluated, our approach allows one to overcome biases in network data, detection algorithms and evaluation metrics by using distinct evaluation strategies when analyzing the quality of such communities. For this, each strategy must strongly highlight a distinct aspect of a community's quality in addition to considering
multiple and diverse detection methods, metrics and network datasets. For example, in Figure~\ref{graphic:ToyNetwork.pdf} the structural and functional aspects of the communities are represented, respectively, by their modularity and similarity with the respective ground truths. Notice that, for our purpose, the choice of the best metrics, detection methods and ground truth data is not important, since we are not trying to identify the best existing community, but the best one among those being compared. 
Thus, when there is a large variety of metrics and methods available for this task, they can be pre-selected based on some criteria of relevance and diversification. 

The rest of this article is organized as follows. Section~\ref{section:fundamentos} reviews related work. Section~\ref{section:approach} describes our approach for community quality evaluation. Then, Section~\ref{section:results} analyzes the experimental results obtained by applying our proposed approach to real and simulated networks. Finally, Section~\ref{section:conclusion} presents our conclusions and some considerations for future work.

\section{Related Work}
\label{section:fundamentos}

\input{methods.tex}

Although community detection has become one of the most popular and best-studied research topics in network science~\cite{Fortunato201075,10.1140/epjds/s13688-020-00223-0,Hric2014,kivela2014multilayer,JCLJISA2018,10.1007/978-3-030-36687-2_22,Zhao2017}, the problem of validating the quality of a community derived from a real network has not received the due attention in the literature, since there is no consensus on what is meant by a good community \cite{YangLeskovec2015}. 
{Moreover, there are several different definitions of what a community is, which has resulted in many distinct approaches to community detection~\cite{10.1145/3341161.3342860,10.1109/TKDE.2019.2911585}.} 
For example, the algorithms listed in Table~\ref{table:communitydetectionalgorithms} usually extract different communities from a given network, which are in general considered of good quality by distinct metrics. 

In a previous work~\cite{JCLJISA2018}, we analyzed the similarity of communities detected by distinct methods in order to identify the set of algorithms that tend to produce more similar results and those that provide more distinct ones, when compared by multiple similarity metrics and considering a given network. 
{Doing so,} we aimed to increase the scope and the consistency of the evaluation process when detecting communities. More recently, Coscia~[2019]\nocite{10.1145/3341161.3342860} proposed an analysis essentially similar to ours also aimed at classifying community detection algorithms according to the similarity of their results. 
{For this, he considers how many times two algorithms provide similar communities.}
{Besides the similarity of the results provided, other criteria have been used to distinguish community detection algorithms.}
For example, Abrahao et al.~[2012], Yang and Leskovec [2015]\nocite{Abrahao:2012:SSC:2339530.2339631,YangLeskovec2015}, and Ghasemian et al.~[2019] \nocite{10.1109/TKDE.2019.2911585} 
consider the structural properties of the detected communities for such a distinction. 

{Although the analysis of the detected communities is commonly used to distinguish the detection algorithms, we have not identified any work that uses the variability of the characteristics of these communities to deal with bias. In fact, in general, such works show some bias when evaluating the quality of a community.} Next, we describe some works that have proposed methods that make it possible to deal with some types of bias 
when assessing the quality of their detected communities. 

\prg{Bias in Community Detection Algorithms.} 
Heuristic algorithms for community detection often find communities that are systematically biased in the sense that their results might be different than the optimum objective function chosen for this specific task~\cite{Abrahao:2012:SSC:2339530.2339631,10.1145/1772690.1772755,Peele1602548}.
Moreover, distinct goals for community detection might lead to totally distinct objective functions~\cite{10.1145/3341161.3342860,10.1140/epjds/s13688-020-00223-0,10.1109/TKDE.2019.2911585}. In addition, the intensity of these differences might vary from a network to another due to detection methods being sensitive to different community structures, topologies, and types or instances of a network~\cite{coscia2011classification,10.1109/TKDE.2019.2911585,10.1145/1772690.1772755}.

Note that detection algorithms are expected to introduce in their final results the same bias of the metrics they use in their optimization functions \cite{JEBABLI2018651,Peele1602548}. A popular case is that of the modularity, conductance and coverage metrics that have strong structural biases that make them favor smaller clusters and whose maximization is aimed at by most detection algorithms~\cite{almeida2012towards,JEBABLI2018651}.

In this context, different approaches have been proposed with the aim of reducing the effect of biases and improving the detection of communities. For instance, Lancichinetti et al. [2012] \nocite{Lancichinetti2012} show how to combine the communities obtained from various detection methods into a consensual one, statistically more stable and with a better structure. It is worth noting that this approach seeks consensus only on the structural aspect of the communities and does not explore the different results produced by multiple methods to identify, analyze and consider any other kind of bias.

\prg{Bias in Evaluation Metrics.}
Existing works usually consider only specific aspects to assess the quality of a community, for example by measuring the structure derived from its connectivity (structural aspect) \cite{NewmanGirvanPhysRevE.69.026113} or by measuring its similarity with a ground truth community (functional aspect) \cite{Peele1602548} to finally performing a comparison with a good baseline score \cite{Hric2014}.
Regarding the structural aspect, community detection algorithms are usually evaluated by correlated metrics or by the same metrics used by their optimization function, such as modularity \cite{Fortunato201075,YangLeskovec2015},
which can produce some biased results. Another type of bias associated with a metric may be an incorrect score systematically attributed according to a specific characteristic present in the data. 
For example, popular quality metrics present strong bias when applied to networks with different sizes or number of clusters~\cite{almeida2012towards,coscia2011classification,Pons2005}. In this context, there is no best metric to assess the quality of a community~\cite{almeida2012towards}.

In particular cases, it is possible to assess the functional aspect of a detected community by comparing it with its respective ground truths~\cite{Hric2014,Peele1602548,zaki2014dataminingbook}. 
For Fortunato et al.~[2010]\nocite{Fortunato201075}, this kind of evaluation involves the definition of a criterion to establish how ``similar'' is a community provided by an algorithm with respect to the ground truth. To address this, the authors adopt some specific metrics such  as \textit{Rand Index} and \textit{Normalized Mutual Information}. 
Note that similarity metrics can provide different scores for a same network \cite{LLD2019SBBD}, in some cases even outliers. In addition, such metrics are also susceptible to producing biased results when comparing communities with specific characteristics \cite{amelio2015normalized,LEI201758}.

To deal with a specific type of bias associated with a metric, works like those carried out by Liu et al. [2019]\nocite{Liu20158693534}, Gösgens et al. [2020]\nocite{gosgens2020systematic}  and Labatut [2015]\nocite{doi:10.1504/IJSNM.2015.069776} usually propose a new metric or enhance an existing one. However, we have not found in the literature any community quality or similarity metric that does not present any bias, i.e., that is capable of dealing with multiple types of bias or does not present any vulnerability to bias originating from other sources, such as detection algorithms or network data. In this sense, our contribution relies on mitigating bias in the evaluation of communities by using multiple existing metrics.

\prg{Bias in Data.}
In addition to the specific biases of each algorithm introduced in the data of the detected communities~\cite{10.1145/1772690.1772755}, there may also be some bias in data from the network~\cite{JEBABLI2018651,JCLJISA2018} and from its ground truth data~\cite{LLD2019SBBD}. In a previous work, Rocha et al. [2017]\nocite{PhysRevE.96.052302} described how the representation of real temporal interactions can result in biased data.  
More recently, Leão et al. [2018]\nocitedb{JCLJISA2018} proposed a solution that avoids the biased data problem by directly removing noise produced by sporadic relationships\footnote{In the history of interactions of a social network there are those that represent a strong relationship between two people in a community (e.g., a teacher and a student in a school) and others, result of chance, that represent interactions between people from different communities and most likely will not occur in the future (e.g., a phone call from a telemarketer)~\cite{LBDL2017SBBD,JCLJISA2018}.} found in a social network. 
They also showed that this kind of noise may cause errors when detecting communities. 
Note that datasets with good ground~truths are rare~\cite{YangLeskovec2015} and might not be always suitable for the type of community we want to analyze~\cite{JCLJISA2018}.

\prg{Final Comments.} In addition to the need of overcoming the specific limitations discussed earlier, some studies report that it is also important to consider multiple strategies when assessing the quality of a community~\cite{dao_bothorel_lenca_2020,JEBABLI2018651,LLD2019SBBD}. In addition, Dao et al.~[2020]\nocite{dao_bothorel_lenca_2020} also systematize a process for obtaining a conclusion on the structural and functional quality of the communities, in this case, by applying the multiple criteria decision making process. 
Jebabli et al.~[2018]\nocite{JEBABLI2018651} also highlight the distinction between different community detection algorithms, however limited to a single comparison criterion that is based on the distribution of the number of communities.

Thus, by analyzing the above works, we identified the following main contributions of our proposed approach to assess the quality of a community:
(i) it deals with different types of bias inherent in data, metrics and algorithms; (ii) it involves different aspects of the quality of a community through the use of multiple strategies; (iii) it uses multiple criteria to distinguish among different community detection algorithms and metrics used to assess them; (iv) it seeks a consensual and consistent decision; (v) it provides requirements for an evaluation maturity model; and, finally, (vi) it provides a framework to systematically assess the community detection task.
These contributions are complementary to our preliminary results presented in our previous work~\cite{LLD2019SBBD}.

\section{Proposed Approach}
\label{section:approach}
\label{subsec:evalImprov}

\begin{figure*}[!t]
     \begin{center}            
\includegraphics[width=1\textwidth]{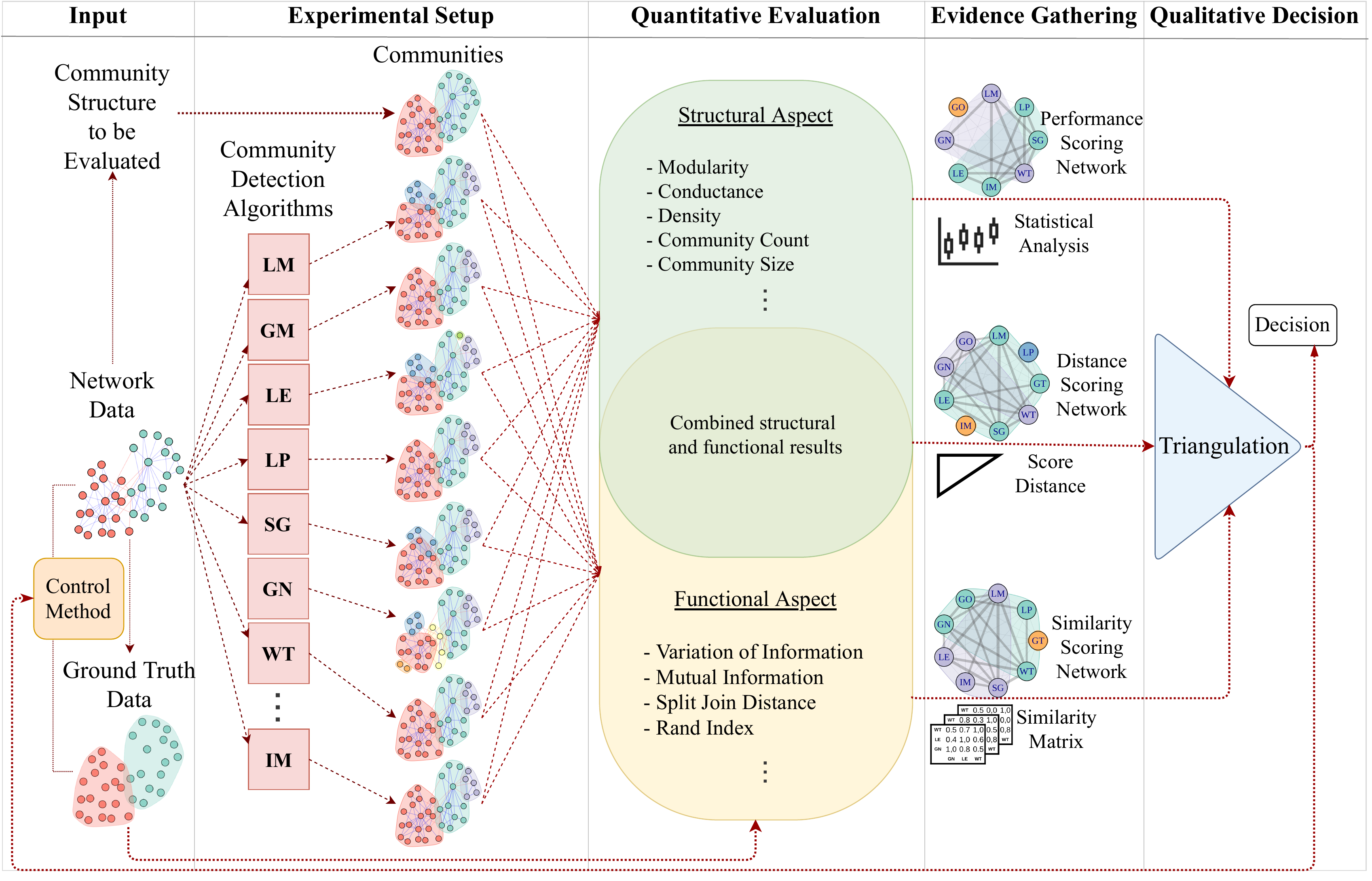}
    \end{center}
    \caption{Overview of the proposed approach to community quality evaluation.}
    \vspace{-2mm}
    \label{fig:approach}
\end{figure*}

Figure~\ref{fig:approach} summarizes our approach for community quality evaluation. First, in the \textit{input} step, we provide a network, its ground truth communities and the set of its communities that we want to evaluate. Next, in the \textit{experimental setup} step, in addition to the set of ground truth communities, we also consider as a further source of evidence the communities detected by distinct algorithms, for example, those listed in Table~\ref{table:communitydetectionalgorithms}. Then, in the \textit{quantitative evaluation} step, all communities are assessed by multiple structural and functional metrics in order to be compared to each other to provide a set of combined evidence, whereas in the \textit{evidence gathering} step we group the results produced by each algorithm in a new set of pieces of evidence to highlight structural and functional aspects related to the quality of each community. Finally, in the \textit{qualitative decision} step, we compare all pieces of evidence to get a final decision on the quality of the communities. However, it is important to notice that in case of a robust decision based on the existing pieces of evidence is not available it is required to apply a control method\footnote{A control method consists in raising a hypothesis about the effect of a given class of entities on an experimental result and testing such a hypothesis by comparing its result with that obtained without considering that class of entities on the data \cite{creswell2018research}.} and start the whole process again to collect additional evidence, as we further describe next.

\subsection{Collecting Evidence} 

By using structural metrics, we are able to quantify the connectivity of specific sets of nodes in the network in terms of structural characteristics that are typical of real-world communities~\cite{YangLeskovec2015}. For this, we take into account multiple pieces of evidence on the quality of a community expressed by the results of statistical analyses of the scores obtained by metrics such as modularity, conductance and density~\cite{NewmanGirvanPhysRevE.69.026113,YangLeskovec2015}. We also use specific statistics, such as the number and size of the detected communities, the distribution of the component sizes, the variance of these values, and other network metrics, such the ones considered in Table~\ref{table:networks_statistics} (see Section 4), to help analyze the results. 

In this step, we also statistically estimate the consensus and the divergence between the detection algorithms with respect to the structure of the communities of a network. For this, we measure the similarity between communities detected by different algorithms, considering similarity metrics such as \textit{Variation of Information}~(VI), \textit{Normalized Mutual Information}~(NMI), \textit{Split Join Distance}~(SJD) and \textit{Rand Index}~(RI)~to provide some functional evidence. By using such additional metrics, we collect evidence on the functional aspects of the detected communities by measuring the similarity between them and their respective ground truths. 

Finally, in the last step (Evidence Gathering), we combine structural and functional aspects in order to make a final decision on the quality of the set of communities provided as input. For this, we measure the structural characteristics of the networks' ground truth communities. Then, we collect evidence about the agreement between such measures and the measures of the communities obtained by distinct community detection algorithms, by using the distance measure between two scores $s_1$ and $s_2$, defined by Equation~\ref{equation:distance}.
\begin{equation}
d(s_1, s_2) = {|s_1 - s_2|}/(s_1 + s_2)     
\label{equation:distance}
\end{equation}
 
\subsection{Diversification Criterion}

In our experimental configuration, we estimate the diversification of community detection algorithms based on the distinction between the communities detected by them and the tendency of pairs of algorithms to detect very different communities. For this, we measure the similarity between the sets of detected communities and the distance $d$ (see Equation \ref{equation:distance}) between the quality score of these communities. 
Then, we model each set of such measures as a multilayer network, by separating in each one of its layers the measures derived from the same metric. In this network, the nodes represent different algorithms and the edges correspond to a relation of similarity among them, weighted by the respective metric value (see Figure~\ref{fig:approach}). Next, a clustering analysis is performed on each layer of the network to identify the algorithms that produce similar results (i.e., belong to the same group). Finally, the algorithms that are often assigned to different groups represent a greater diversification in the baseline repertoire.

Regarding this multilayer network, we also analyze the diversification of the quality measures. For this, we measure the similarity between different layers. Thus, each pair of layers with very similar clusters reinforces the respective pairs of evidence collected in those layers,  when they correspond to uncorrelated metrics. On the other hand, very different clusters indicate diversity of measurements and can be an evidence of the effect of bias in some of the metrics. Then, this set of evidence is considered to decide on the final quality of the communities.

\subsection{Decision by Triangulation}

By analyzing all pieces of evidence considered (structural, functional and the two combined), we capture distinct aspects of the communities' quality and conclude on the quality of their structures by means of a consensual decision.
For this, we first cross-check these pieces of evidence to estimate their validity and consistency. Thus, any agreed evidence on a specific aspect reinforces the internal consistency of that strategy on that specific aspect. On the other hand, the agreement on pieces of evidence collected considering different aspects allows our approach to validate the quality evidenced by the corresponding strategies.

The final decision on the quality of the community structures is obtained when a consensual consistency among all strategies is achieved.
Possibly, the disagreement between different pieces of evidence explains the internal bias of a strategy (for example, a disagreement in one of its metrics) or even a bias of the strategy itself (when a set of its pieces of evidence disagrees with a set of pieces of evidence produced by another strategy).
In this analysis, it is also carried out a cross-checking of the consistency and validity of the pieces of evidence with respect to the quality of the communities detected by a specific algorithm.

\subsection{Controlling Data Bias}

To strengthen the pieces of evidence on the quality of a community and allow a more consistent and consensual conclusion, 
we first verify the existence of bias in the network data and in its ground truth data. Then, we check the effect of data bias by controlling its source. More specifically, we estimate and minimize the effect of bias by removing from the network nodes and edges that systematically damage its structure. Note that here ``data bias'' is any error generated by a community detection algorithm that might be associated with some noise in the network being assessed.

For example, after collecting evidence about the structural and functional quality of the communities existing in a network, a low consensus among very convincing measures that express such a quality is an indication that the source of bias is the data. To verify this, we apply a control method, i.e., we deliberately remove from the network structure the part supposedly biased by the use of a specific network filter. Then, we apply the entire evaluation flow shown in Figure \ref{fig:approach} and collect additional pieces of evidence for each evaluation strategy. If this procedure leads to a higher level of consensus (not necessarily indicating higher quality communities), we obtain the confirmation that the structure removed from the data influenced the assessment, possibly in a systematic way. 
Note that we adopt three different types of control filter: for noisy edges, noisy nodes and small components, which are described next.

To filter out noisy edges, we use the framework proposed in our previous work~\cite{JCLJISA2018}. When identifying nodes in the network that belong to an entity class that violates the community structure, we use a filter per class, i.e., we remove from the network (and from its ground truth) all nodes labeled in the respective ground truth with that class identifier. With the removal of these noisy nodes and edges, we aim to extract communities with a better defined structure and, most importantly, that allow a more objective assessment and the collection of evidence to decide on the quality of the communities. Finally, the third filter, used as a control method, aims to remove very small and dense components from the network, since they contribute little to distinguish specific bias from the algorithms, but can distort the structure of the detected communities and influence their quality assessment due to the inherent bias of algorithms and metrics.

\section{Experimental Results}
\label{sec:results}
\label{section:results}

To evaluate our proposed approach, we run a series of experiments to assess the communities derived from twelve networks by applying a combination of eight algorithms based on state-of-the-art community detection methods. 
Note that in these experiments we analyze the communities generated by each algorithm separately, considering the other ones as their baselines. In addition, experiments involving non-deterministic algorithms (see Table~\ref{table:communitydetectionalgorithms}) were performed several times (at least 30 repetitions) to ensure the reliability of the results.

\input{networks.tex}

\subsection{Networks}

Initially, we modeled as temporal and aggregate edge graphs the following scientific collaboration networks (here identified according to their respective datasets): 
APS - coauthorship network of members of the American Physical Society ~\cite{Brandao2017,JCLJISA2018}; 
PubMed - coauthorship network derived from scientific articles available in MEDLINE~\cite{Brandao2017,LLD2019SBBD}; 
arXiv - coauthorship network derived from scientific articles deposited in arXiv~\cite{JCLLD2018};
SIC10 - coauthorship network of papers presented at the 2019 Seminar on Scientific Initiation held at the Federal Institute of the Northern of Minas Gerais~\cite{RosaAbdiel2019SIC10anosXISimposioJanuaria}; 
LH10 - contact network between people in a hospital~\cite{10.1371/journal.pone.0073970}, 
HSC - contact network in a high school~\cite{Gemmetto2014,Genois2018}; 
PSC - contact network in a primary school~\cite{10.1371/journal.pone.0023176}; 
IVS13 and IVS15 - contact networks in a French health institute in two different years~\cite{genois_vestergaard_fournet_panisson_bonmarin_barrat_2015,Genois2018};
ACM09 - contact network in the ACM 2009 Hypertext Conference~\cite{ISELLA2011166};
HSC(S) - simulated network based on the HSC network~\cite{JCLLD2018}; 
EEU(S) - simulated network based on an e-mail exchange network~\cite{JCLLD2018}.

In addition, we modeled the metadata of these networks to use them as ground truths for their respective communities, where each community is identified by the predominant research area of their respective researchers in the collaboration networks, by the class of the students in the high/primary school contact networks, and by the department of the workers in the the French Health Institute contact  networks. With respect to the Hypertext ACM Conference network, we notice that it represents a single community. 
Finally, the ground truths of the simulated networks are the same of the real networks from which they were generated \cite{grm}.

Table~\ref{table:networks_statistics} presents a general characterization of these networks. Note that, it was important to consider in our experiment synthetic and real-world networks that have many variations of type, size and complex structural features. In addition, the number of networks used is justified by the need to collect sufficient evidence to obtain the consistent results.

\begin{figure}[ht]
    \centering
    \includegraphics[width=0.497\textwidth]{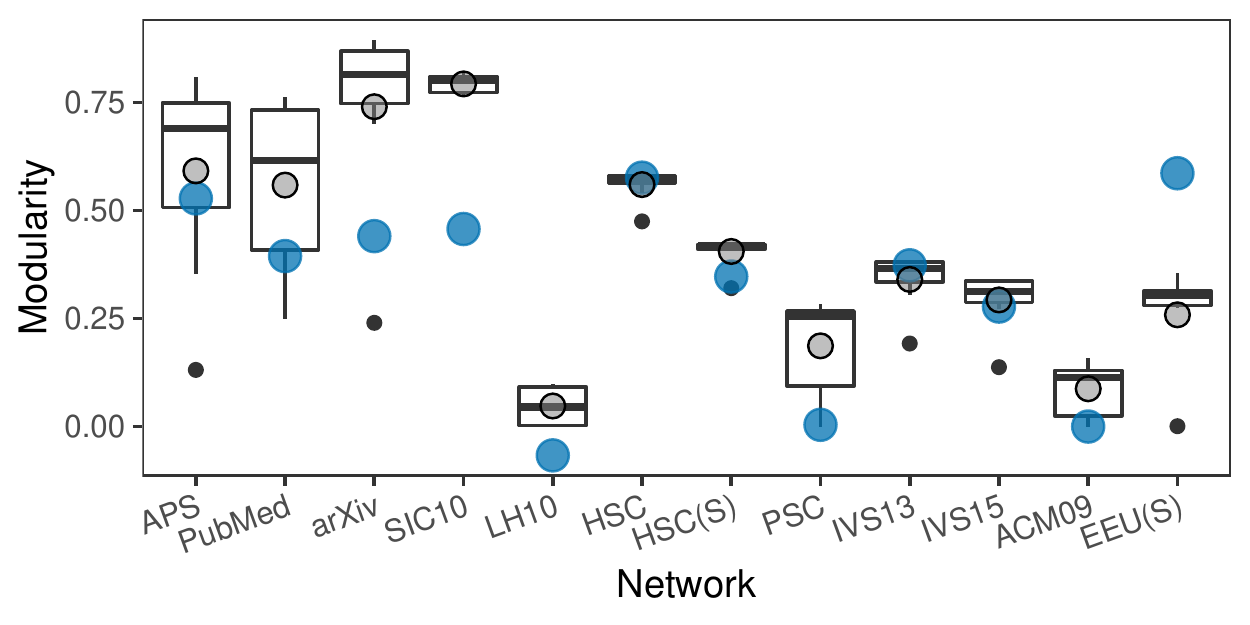}    
    \includegraphics[width=0.497\textwidth]{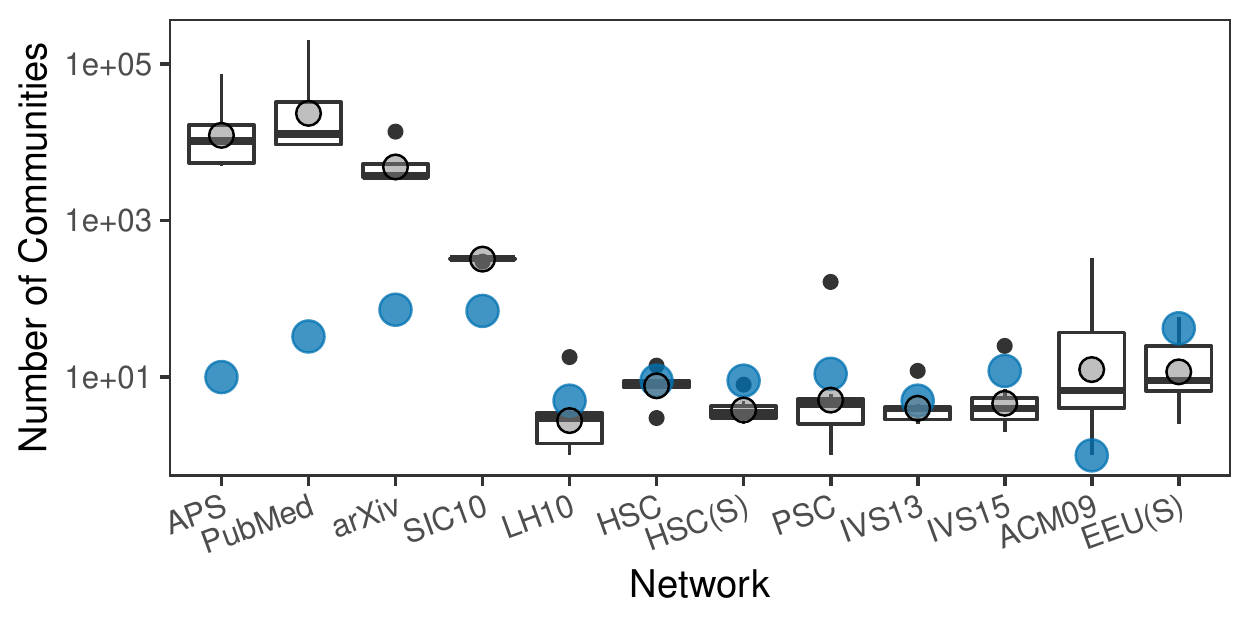}
    \caption{{Details of some of the pieces of evidence considered in the structural evaluation strategy:} (left) modularity values for the communities detected by all considered algorithms in each network (boxplot) and for the respective ground truths (blue dot); (right) number  of communities detected by all considered algorithms in each network (boxplot) and in the respective ground truths (blue dot).}
    \label{graphic:structuralevaluation}    
\end{figure}

\subsection{Evidence Considered}

The combination of functional and structural evidence in our experiments allowed us to corroborate the quality of the ground truths as well as of the communities detected in all networks. This also made it possible to indicate the algorithm that identified the best communities in the networks. For this, we first analyzed the results of each strategy individually, providing hypotheses about the quality of the communities. Then, we combined these results, verifying the consensus among the quality of the communities. 
In this way, we verified which hypotheses were refuted, as well as the biases identified. 

\subsubsection{Identifying the Best Communities}

\begin{figure}[ht]
\includegraphics[width=0.497\textwidth]{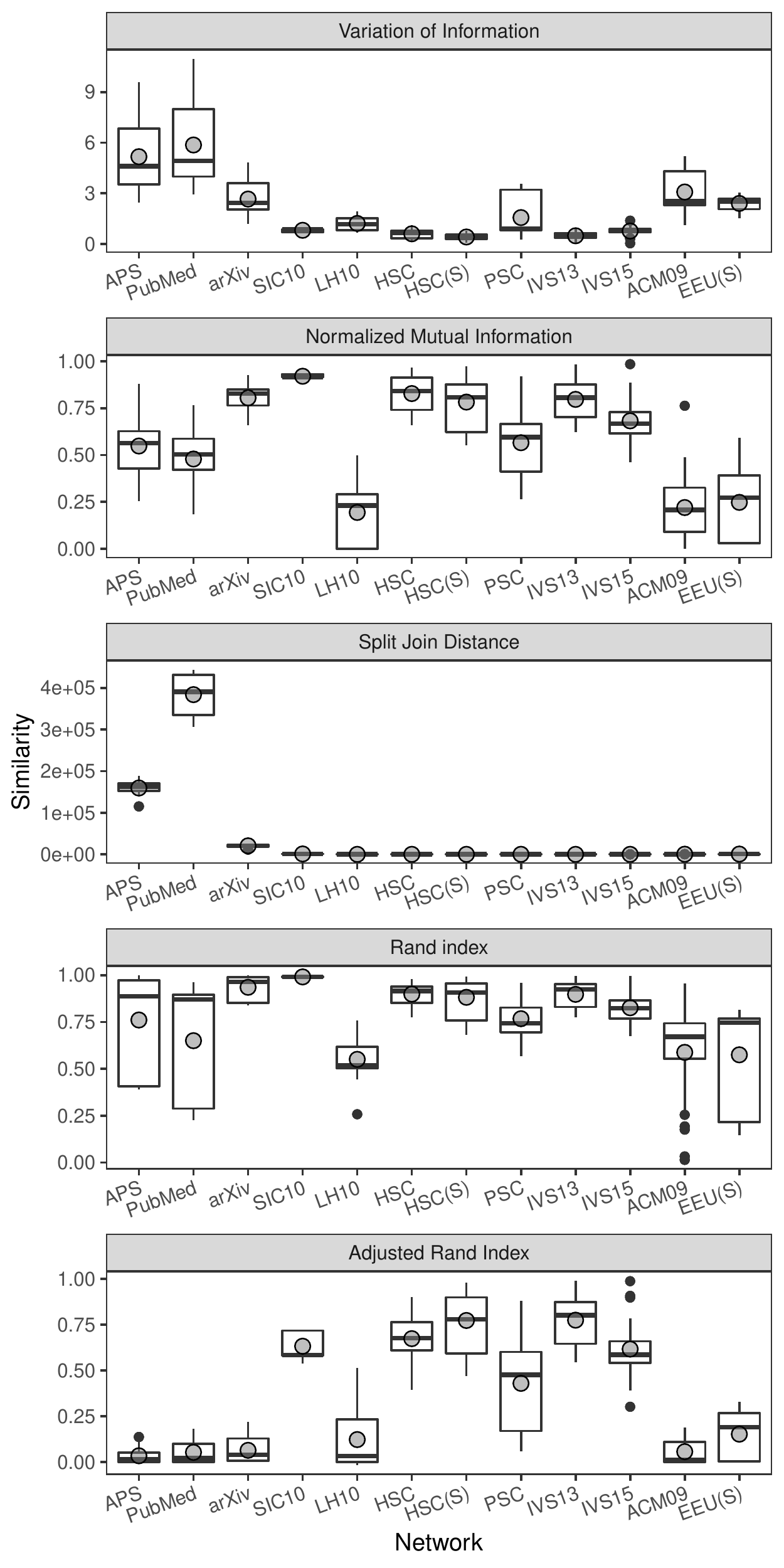}
\includegraphics[width=0.497\textwidth]{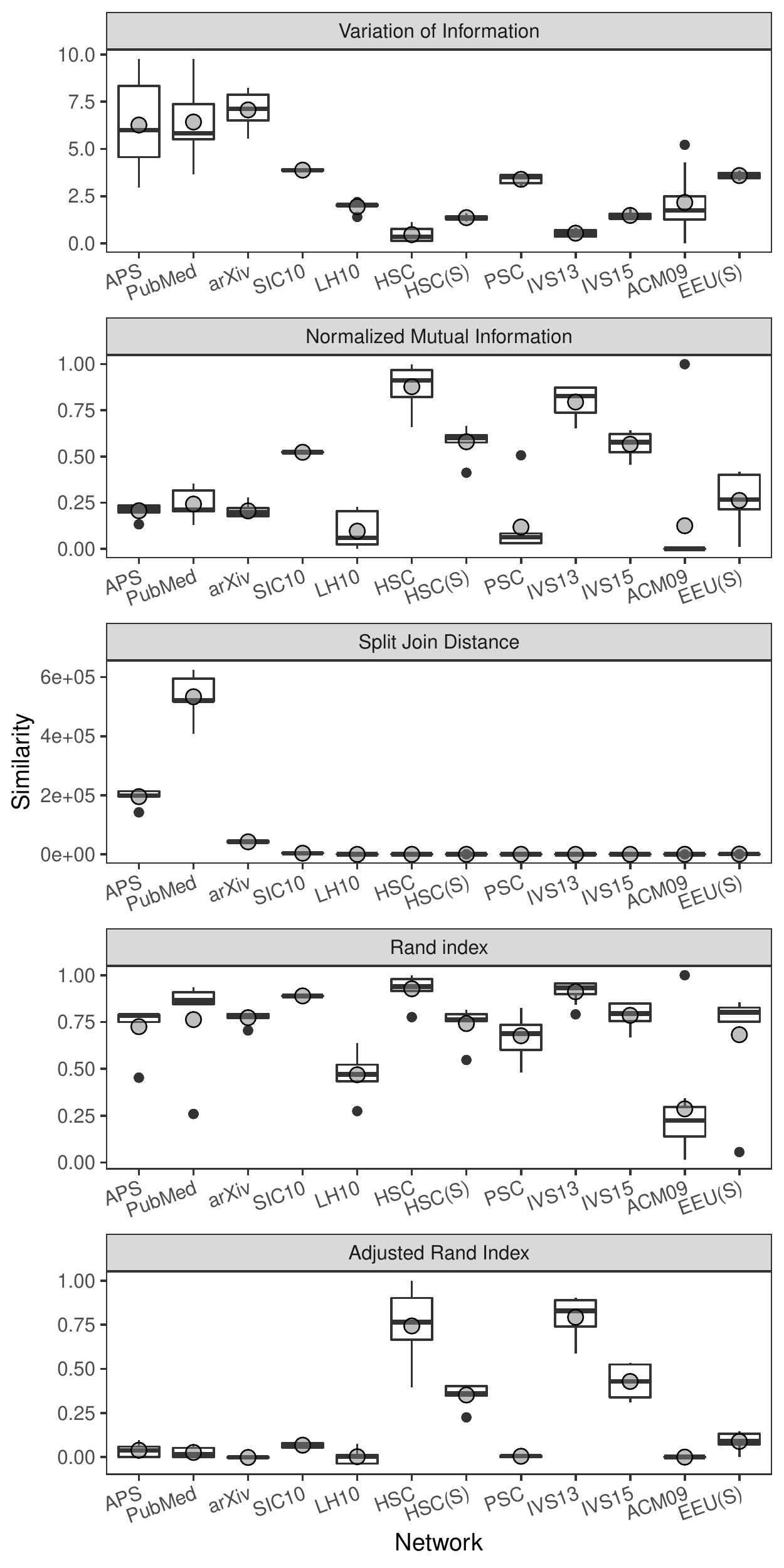}
\caption{Details of some of the pieces of evidence considered in the functional evaluation strategy: (left) similarity values between the communities detected by different algorithms and expressed by different metrics; (right) similarity values between the ground truth and the communities detected by distinct algorithms and expressed by different metrics. Note that, especially for the metrics VI and SJD, the lower their values, the greater the similarity indicated.}
\label{graphic:functionalevaluation}
\end{figure}

First, we analyze the structure of the communities detected by the different algorithms. Here, we note that the communities derived from the HSC, SIC10 and arXiv networks present the best defined characteristics. 
For this, we considered the following pieces of evidence:
high average modularity (Figure~\ref{graphic:structuralevaluation}), greater consensus on the structure of the communities (interquartile of the similarity between them, presented in Figure~\ref{graphic:functionalevaluation}), greater confidence of the modularity  value obtained in different experiments with the same non-deterministic algorithm (coefficient of variation less than 0.1 of the modularity values between repetitions of detection experiments) and small variation in the number of communities detected by
these algorithms (Figure~\ref{graphic:structuralevaluation}). However, as we shall see below, although such pieces of evidence indicate that the communities from these three networks have the same characteristics, we have not come to the same conclusion about their quality. 

From a functional viewpoint, unlike the High School network, in the arXiv and SIC10 networks there is no convergence of evidence to confirm the quality of their communities when compared with their ground truths (Figure~\ref{graphic:functionalevaluation} (right)). 
This can be considered as a disagreement with respect to the structural aspect when we compare the distance between the structural measure values, such as modularity values or number of communities of the arXiv and SIC10 networks with those of most other networks. 

Note that in Figure~\ref{graphic:structuralevaluation} (left), for example, there is a large difference between the modularity values of the ground truths and those estimated for the detected communities in the arXiv and SIC10 networks. In addition, according to Figure~\ref{graphic:structuralevaluation}, the number of communities in the ground truths is far from the number of communities actually detected in the networks\footnote{The distance between these two values (number of communities) is measured by the metric given by Equation 1.}. Therefore, the strength of these initial pieces of evidence has led us to the conviction that the communities detected in the actual networks are the correct ones. In addition, the confidence intervals of the measures and the structural evidence that strongly disagree with the functional one corroborate the interpretation that the detected communities are the real ones and not those shown by the ground truths. 
This means that the two sets of evidence, structural and functional, contradict each other on which communities in the arXiv and SIC10 networks are the best ones. Thus, we need to make a decision on which set of evidence is the strongest one, the structural quality of the ground truth or the detected communities. Based on our approach, the possibility of bias being the cause of these divergences makes it necessary to evaluate them by using a third set of evidence (on a new and independent particular aspect) to support one of the two contradictory sets of evidence. To do so, we have analyzed these contradictory sets of evidence in order to raise some hypotheses about the main source of bias in the convergence of the results  on the quality of the communities of the arXiv and SIC10 networks.

As we can see, in the arXiv and SIC10 networks the bias caused by the detection algorithms does not considerably interfere in their results, since the communities suggested by them are structurally similar. In addition, this was evidenced in these networks by all structural metrics considered, whose values corroborate a high-quality community structure, as already shown in Figures~\ref{graphic:structuralevaluation}  and~\ref{graphic:functionalevaluation}. Thus, we have hypothesized that the bias interference is predominantly in the data, provoking a disagreement between pieces of structural and functional evidence, as well as the identification of false communities of high quality.

For this, we first analyze the ground truth communities of the scientific collaboration networks and then consider the meaning of these communities in those networks, i.e., they are groups of researchers that publish together and predominantly in the same area of knowledge. However, it should be noted that this definition is not absolute, since there may be a multidisciplinary community with sporadic co-authorships or a community of researchers that work in the same area, but do not collaborate with each other. In both possible cases, ground truth communities are not very well captured by detection algorithms that rely on network connectivity.
Thus, we consider the hypothesis that the bias that obscures the real community structure of the arXiv and SIC10 networks is a consequence of the existence of edges and nodes that represent, respectively, sporadic collaborations and researchers that work in the same knowledge area, but do not significantly interact with each other. 
Then, to test our hypothesis, we run the following bias control experiment: we removed such skewed edges and nodes using the filtering framework proposed by Leão et al. [2018]\nocite{JCLJISA2018}, and then, as a new iteration, followed again the steps of our evaluation approach as shown in Figure~\ref{fig:approach}, using as input the filtered version of that network\footnote{The generated datasets are available by request at {http://cnet.jcloud.net.br} repository.}.
From this new iteration, we obtained new communities in which the structural and functional metrics of the arXiv network converged, as indicated by its greater similarity with the ground truth communities, and showing a better structural aspect, as indicated by all metrics used for this purpose (Figure~\ref{graphic:structuralevaluation}).

However, this same convergence was not observed in networks such as SIC10, which, therefore, was also inspected by a new iteration of the evaluation flow of our approach. This time, the second control method used was the filtering of the smallest connected components, due to the considerable number of them in this network. As a result, it was found that the measurement of the structural quality of their communities is biased by the presence of these components. Finally, with these findings, a consensual decision was reached among the evaluation strategies, thus confirming that the communities of both networks are of \textit{High Quality}. Note that the components that distorted the structural quality of both networks to \textit{Very High} correspond only to noise in the SIC10 network (5\% of its structure) and in the arXiv network itself (90\% of its structure).

\input{table/bestcommunitiessummary.tex}

This way, we have been able to identify that the most significant source of bias in the arXiv network was in its data and in the SIC10 network in the structural metrics used, such as modularity. This influenced the premature results of assessing the quality of the communities in these networks, respectively underestimating and overestimating their values. Furthermore, these two situations exemplify how considering few pieces of evidence can lead to apparently very convincing results, but unreliable. 
Note also that the divergence between consensus conclusions obtained by different strategies is somewhat recurrent in the premature conclusions of evaluating communities in all the networks used. More precisely, in 60\% of the networks, the quality measured by the structural or functional strategies showed underestimated values and in 30\% of the networks, they presented overestimated values in relation to the decision considering the influence of biases. Table III shows all premature decisions about the quality of communities, as well as the consensual decision obtained by identifying and considering biases. It is worth highlighting some of these cases, as follows.

In the APS and PubMed networks, data bias is also the main source of divergence between sets of evidence, beyond bias in structural metrics and detection algorithms.  
The communities of the LH10, ACM09 and PSC networks presented themselves with very low quality, indicated individually by the scores in all strategies. Despite this, the low confidence of these estimates (also widely evidenced as detailed in Figures.~\ref{graphic:structuralevaluation} and~\ref{graphic:functionalevaluation}) made us investigate and check for bias in their data, in part, caused by the presence of noisy edges, as defined by Leão et al.~[2018]. In addition, in the LH10 network data, we verified a class of nodes that tend to violate the community structure, and that, when isolated, allowed to test its effect on the evaluation result. Finally, our assessment approach revealed communities with different quality factors in these networks: from low to very high. The PSC network stands out with communities with very high quality revealed by our approach and which demonstrates that the premature decision was very wrong and influenced by the noisy edges. Highlighting milder mistakes in their premature decisions, the identification and isolation of biases in the EEU(S), HSC, InVS15, HSC(S) and InVS13 networks, allowed us to conclude that these networks have high quality communities. In general, the final decision on the quality of the networks was obtained with two iterations of our assessment approach, the first always made on the data of the skewed network and the other controlling the bias of these data by filtering noisy edges (except on LH10 and SIC10, which required a third iteration and specific control methods).

\subsubsection{Best Detection Algorithms}

Although the most modular communities are those detected by the Louvain algorithm (upper bounds shown in Figure~\ref{graphic:structuralevaluation}, left), the modularity values of the communities detected by the Infomap algorithm are generally closer to those of the ground truth (there is a greater agreement between them). In addition, Infomap provided several cases in which there was an agreement between the modularity of the detected communities and that of the ground truth. On the other hand, these same metrics achieved smaller values for the Louvain algorithm. Moreover, the modularity of the communities extracted by different algorithms and that of the functional communities have considerably varied for most networks. We also verified how different community detection algorithms agree with each other and with the network ground truths with respect to their communities. Despite the variation in the structure of the detected communities, as shown in Figure~\ref{graphic:functionalevaluation} (left graph), there was a higher consensus among them than with respect to their ground truth communities (Figure~\ref{graphic:functionalevaluation} (right graph)). 
\vspace{-5mm}
\input{table/bestalg.tex}
\vspace{-2mm}

In addition to obtaining a consensus among different algorithms, our approach also identified some algorithms with distinct behavior, such as Infomap, that detected less modular communities, but in general more similar to their ground truths. Despite such divergences among the strategies, most pieces of evidence indicate the Louvain algorithm as the least biased among those that are based on the modularity maximization and the one that obtained estimates with higher values for most of the structural metrics, particularly modularity. 
We also identified some algorithms that presented the best score on some specific metrics. This is the case of the Louvain  algorithm (LM) for modularity, and of the Infomap (IM) and Leading Eigenvector (LE) algorithms for the similarity metrics Normalized Mutual Information (NMI), and Split Join Distance (SJD) and Variation of Information (VI), respectively (see Table~\ref{table:bestmethod}).
Notice that our proposed approach is able to analyze distinct alternative solutions for the task at the hand, thus being able to identify those algorithms that provide the best trade-off.  

\section{Conclusions and Future Work}
\label{section:conclusion}
The main contribution of this paper is an approach to identify and reduce the effect of biases when assessing the quality of a set of communities. 
Specifically, we use multiple and diversified measurement strategies designed to capture different aspects of the quality of a community structure. For its evaluation, we carried out a set of experiments using twelve networks (ten real ones and two synthetic ones) and compared the results obtained by eight community detection algorithms considered the state-of-the-art in the area. In addition, we also used distinct metrics, each one providing a piece of evidence from a specific point of view on the quality of the assessed communities.

In this context, we consider the consensus and the divergence between such pieces of evidence to hypothesize about the influence of bias coming from metrics, algorithms or network data when assessing the quality of such communities. Thus, outliers observed by statistically analyzing the resulting measurements allowed us to test hypotheses of specific biases with respect to metrics and detection algorithms. To test a hypothesis on bias in network data or in its ground truth metadata, we use control methods, such as node filters or network edges. These methods allowed us to verify that, by removing the supposed biased structures from the data, a greater consensus in the evaluation of their communities can be verified by the multiple metrics and strategies used by our approach.

By doing so, we were able to sustain our hypothesis by showing that the quality evaluation of communities detected from a network must be supported by multiple pieces of evidence. That is, given the discrepancy between the quality indicated by distinct evaluation strategies, we evidentiate that the use of a single quality metric or strategy, be it structural or functional, makes the results biased and unreliable. On the other hand, our multi-strategy evaluation approach makes it possible to explain extreme values for some of the metrics considered and decide which strategies lead to a more consistent conclusion about the quality of a community. For example, we were able to verify the existence of bias in some metrics, network data and detection algorithms, which allowed us to reach a consensual decision very different from the premature decision, individually suggested by the metrics used by one of the strategies. 

A current limitation of our proposed approach is the use of a predefined set of evaluation metrics and community detection algorithms. However, this limitation can be easily overcome by providing a configurable framework in which such features could be defined according to specific characteristics of the networks being considered. It is also worth noting that the approach proposed in this article can be applied to other types of algorithm  (such as those for clustering tabular data, backbone extraction, core-periphery analysis, detection of dynamic or overlapping communities, etc.), as well as adapted to other tasks besides community detection (such as system modeling or simulation, supervised machine learning techniques, missing data prediction, etc.). 
Finally, another line of future work could be, for example, adapting this approach to assess the task of link prediction in social networks in order to provide more robust results.

\section*{Acknowledgements}

Work supported by project MASWeb (FAPEMIG/PRONEX grant APQ-01400-14) and by the authors' individual grants from CNPq and FAPEMIG. Particularly, the first author would like to thank LBD/UFMG, JCLoud.net.br and LabSiCCx - Laboratório de Sistemas Computacionais Complexos (PROPPI/IFNMG, project Nr. 209/2019) for the infrastructure provided.

%% file: methods.tex
\begin{table}[!tb]
\begin{center}
\caption{{Methods for community detection.}}
\vspace{4mm}
\label{table:communitydetectionalgorithms}
\resizebox{0.8\textwidth}{!}{
\begin{threeparttable}
    {
\begin{tabular}{@{}llcl@{}}
\toprule
Main Method & Algorithm & $\xi$ & References                                \\ \midrule\midrule
\multirow{3}{*}{\begin{tabular}[c]{@{}l@{}}Modularity\\maximization\end{tabular}}              & Louvain Modularity (LM)                                         & D     & \cite{blondel20081742-5468-2008-10-P10008} \\ 
                                                      & \begin{tabular}[c]{@{}l@{}}Greedy Optimization\\ of Modularity (GM)\end{tabular} & D     & \cite{Clauset2004PhysRevE.70.066111}       \\ 
                                                      & Leading Eigenvector (LE)                                        & D     & \cite{Newman2006}                          \\ \midrule
\multirow{2}{*}{\begin{tabular}[c]{@{}l@{}}\REVIEWED{Dynamic process}\end{tabular}}                                 & Label Propagation (LP)                                          & N     & \cite{Raghavan2007}                        \\
& {Spin-glass (SG)}                                          & N     & \cite{PhysRevE.74.016110} \\
\midrule
\begin{tabular}[c]{@{}l@{}}Removal of edges\\ between communities\end{tabular}                  & Girvan–Newman (GN)                                           & D     & \cite{NewmanGirvanPhysRevE.69.026113}      \\ \midrule
\multirow{2}{*}{\begin{tabular}[c]{@{}l@{}}Node closeness given\\by random walks\end{tabular}} & Walktrap (WT)                                                   & N     & \cite{Pons2005}                            \\
                                                      & Infomap (IM)                                                    & N     & \cite{10.1371/journal.pone.0018209}        \\ \bottomrule
\end{tabular}
\par
\begin{tablenotes}
\item $\xi$: State model (D-Deterministic/N-Non deterministic).
\vspace{-2mm}
\end{tablenotes}
}
\end{threeparttable}}
\end{center}
\end{table}

%% file: networks.tex
\ifx\NOSHOWFIG\defined
\begin{table}
\centering
\caption{{Characterization of the networks.}}
\vspace{4mm}
\label{table:networks_statistics}

\resizebox{1\textwidth}{!}{
\begin{threeparttable}
    {
\begin{tabular}{@{}llrrrrrrr@{}} 
\toprule
Application Domain & \multicolumn{1}{c}{{Network}} & \multicolumn{1}{c}{$|V|$} & \multicolumn{1}{c}{$|E|$} & \multicolumn{1}{c}{$\Delta$} & \multicolumn{1}{c}{$D$} & \multicolumn{1}{c}{\emph{CC}} & \multicolumn{1}{c}{\emph{C}} \\ \midrule\midrule
\multirow{3}{*}{\begin{tabular}[l]{@{}l@{}}Scientific  Collaboration\end{tabular}} & \iAPS~\cite{Brandao2017}  & 181k & 852k  & 305 & 0.5 & 0.33 & 5k \\ 
 & \iPUBMED~\cite{Brandao2017} & 444k & 5.5M & 4869 & 0.6 & 0.36 & 9k\\ 
 & {arXiv}~\cite{JCLLD2018} & 33k & 180k & 424 & 3.3 & - & 3k \\
 & {SIC10}~\cite{RosaAbdiel2019SIC10anosXISimposioJanuaria} & 3k & 12k & 96 & 20 & 0.34 & 119 \\
\hline
\begin{tabular}[l]{@{}l@{}}Contact in a Hospital\end{tabular} & {LH10} ~\cite{10.1371/journal.pone.0073970} & 76 &  1k & 65 & 4k & 0.6 & 1 \\
\hline
\multirow{2}{*}{\begin{tabular}[l]{@{}l@{}}Contact~in~a~High~and \\Primary~School\end{tabular}} & {HSC}~\cite{Genois2018} & 327 &  5818 & 87 & 1k & 0.44 & 1\\
 & {PSC}~\cite{10.1371/journal.pone.0023176} & 242 & 8k &  134 & 3k & 0.48 & 1 \\
\hline
\multirow{2}{*}{\begin{tabular}[l]{@{}l@{}}Contact in a French Health \\ Institute \end{tabular}} & {IVS13}~\cite{genois_vestergaard_fournet_panisson_bonmarin_barrat_2015} & 92 & 1k & 44 & 2k & 0.37 & 1 \\
& {IVS15}~\cite{Genois2018} & 217 & 4k & 84 & 2k & 0.36 & 1 \\
\hline
\begin{tabular}[l]{@{}l@{}}Contact in the Hypertext\\ACM Conference\end{tabular} & {ACM09}~\cite{ISELLA2011166} & 403 & 10k & 189 & 1k & 0.24 & 1 \\
\hline
\multirow{2}{*}{\begin{tabular}[l]{@{}l@{}}Simulated Networks\end{tabular}} & {HSC(S)}~\cite{LLD2019SBBD} & $\approx$327 & $\approx$7k & $\approx$116 & $\approx$1k & $\approx$0.4 & 1 \\
& {EEU(S)}~\cite{JCLLD2018} & $\approx$1k & $\approx$10k & $\approx$78 & $\approx$267 & $\approx$0.29 & 2\\
\bottomrule
\end{tabular}
\par
\begin{tablenotes}
\small{\item $|V|$: set of vertices; $|E|$: set of edges; $\Delta$: max degree; $D$: density (x${10^{-4}}$); $CC$: cluster coefficient; $C$: number of components. The min degree is 1 in all networks. 
}
\end{tablenotes}
}\end{threeparttable}}
\end{table}

%% file: table/bestcommunitiessummary.tex
\begin{table}[ht]
\centering
\caption{Decision on the quality of the communities.}
\label{box:decision}
\resizebox{\textwidth}{!}{%
\begin{threeparttable}
    {
\begin{tabular}{l|ll|ll|c|l|l}
\hline
\multicolumn{1}{c|}{\multirow{2}{*}{Network}} &
  \multicolumn{2}{c|}{Premature Decision} &
  \multicolumn{2}{c|}{Data Quality} &
  \multicolumn{1}{c|}{\multirow{2}{*}{\begin{tabular}[c]{@{}c@{}}Control\\ Method\end{tabular}}} &
  \multicolumn{1}{c|}{\multirow{2}{*}{\begin{tabular}[c]{@{}c@{}}Main Source of\\ Significant Bias\end{tabular}}} &
  \multicolumn{1}{c}{\multirow{2}{*}{\begin{tabular}[c]{@{}c@{}}Consensual\\ Decision\end{tabular}}} \\ \cline{2-5}
\multicolumn{1}{c|}{} &
  \multicolumn{1}{c|}{Structural} &
  \multicolumn{1}{c|}{Functional} &
  \multicolumn{1}{c|}{Net} &
  \multicolumn{1}{c|}{Ground Truth} &
  \multicolumn{1}{c|}{} &
  \multicolumn{1}{c|}{} &
  \multicolumn{1}{c}{} \\ \hline\hline
LH10   & Very Low  & Very Low       & Low      & Medium    & NF,EF & Net and GT    & Low \\ \hline
PubMed & Medium    & Low       & Medium   & Very Low  & EF & GT, Alg and Met & Low \\ \hline
APS    & High      & Low       & Medium   & Low       & EF & GT, Alg and Met & Low \\ \hline
ACM09  & Very Low  & Very Low  & Low      & Very Low  & EF & Net and GT    & Medium \\ \hline
EEU(S) & Low       & Medium    & Very Low & High      & EF & Net           & High \\ \hline
arXiv  & Very High & Low       & Medium   & Low       & EF & Net and GT    & High \\ \hline
SIC10  & Very High & Medium    & High     & Medium    & EF,CF & GT and Met    & High \\ \hline
PSC    & Very Low  & Very Low  & Low      & Medium    & EF & Net           & Very High \\ \hline
HSC(S) & Low       & High      & Medium   & Very High & EF & Net           & Very High \\ \hline
InVS15 & Low       & High      & Medium   & Very High & EF & Net           & Very High \\ \hline
InVS13 & Low       & High      & Medium   & Very High & EF & Net           & Very High \\ \hline
HSC    & High      & Very High & High     & Very High & EF & None      & Very High \\ \hline
\end{tabular}%
\begin{tablenotes}
\small
\item Main source of significant bias: GT (ground truth data); Alg (algorithms); Met (metrics); Net (network data). Control method: NF (node filter);  EF (edge filter); CF (component filter).
\end{tablenotes}
}
\end{threeparttable}
}
\end{table}

%% file: table/bestalg.tex
\begin{table}[ht]
\caption{Best detection algorithms according to distinct experiments.}
\begin{center}
\resizebox{1\textwidth}{!}{
\begin{threeparttable}
    {
\begin{tabular}{@{}llllllll@{}}
\toprule
Best Algorithm  & LM            & GM  & LE          & LP             & WT               & IM                    \\ \midrule\midrule
Metrics & ARI, SJD      & ARI & SJD*, VI*     & RI, SJD, VI    & ARI, RI          & ARI, NMI*, RI, SJD, VI\\ 
Networks & PubMed, arXiv & APS & APS, PubMed & APS, Sinth. & arXiv, Sinth. & All                  
\\ \bottomrule
\end{tabular}
\par
\begin{tablenotes}
\item *Metrics with the best overall value.
\end{tablenotes}
}
\end{threeparttable}}
\end{center}
\label{table:bestmethod}
\end{table}